\documentclass[12pt]{iopart}
\usepackage{iopams}  
\usepackage{graphicx,subfigure}
\usepackage{ulem}
\usepackage{color}

\begin{document}

\title {Bloch oscillations in supersolids}

\author{Muhammad.~S.~Hasan$^1$, J.~Polo$^{1,2}$, J.C.~Pelayo$^1$ and Th.~Busch$^1$}

\address{$^1$Quantum Systems Unit, Okinawa Institute of Science and Technology Graduate University, Okinawa, Japan 904-0495}
\address{$^2$Quantum Research Centre, Technology Innovation Institute, Abu Dhabi, UAE}

\ead{muhammad.hasan@oist.jp}

\vspace{10pt}

\begin{abstract}
We show that the motion of an accelerated atomic impurity immersed in a spin-orbit coupled Bose-Einstein condensate in the supersolid stripe phase undergoes oscillations, similar to the well-known phenomenon of Bloch oscillations in solids. While the back-action of the oscillatory movement onto the condensate excites phonon modes inside the supersolid, 
it does not affect the position of the roton minimum and therefore not the periodicity of the matter wave lattice. The ultimate decay of the oscillations is mostly due to the dispersion of the wavepacket and we show that this can be counteracted to a large extent by assuming that the impurity is a bright soliton.
\end{abstract}

\section{Introduction}
Ultracold atoms in optical lattices have, in the last two decades, been fruitful systems to study the physics of periodic quantum many-body systems in clean and highly controllable settings \cite{Bloch:08,Gross:17}. These range from the paradigmatic realisation of the bosonic \cite{Greiner:2002} and fermionic \cite{Jordens:2008,Schneider:2008} Mott insulator transitions, to the creation of systems with topological quantum matter \cite{Goldman:2016}. However, as optical lattices are static potentials, they do not allow for phonon modes and therefore any possible connection to condensed matter systems is incomplete. While Hubbard models with phonons have been suggested to be in principle realisable in systems made from self-assembled crystals of polar molecules \cite{Pupillo:2008}, these have not been experimentally observed yet. At the same time, the creation of optical lattices that are capable of sustaining phonon modes is a highly active topic as well \cite{guo2021optical}.

Another way to mimic lattice phonons is to engineer atomic mixtures, where one species is trapped in an optical lattice, while spatially overlapping with a second species that is Bose condensed. Scattering between these two components would allow the atoms trapped in the lattice to effectively couple to the Bogoliubov excitations of the Bose-Einstein condensate (BEC), which play the role of phononic degrees of freedom \cite{parish:2015,parish:2018}. However, this is a very complex setup and in this work we consider a significant simplification by replacing the periodic optical potential by a standing matter wave formed by the BEC itself. 

Periodic matter waves can be created through interference in BECs \cite{Andrews:1997} and scattering from standing matter waves has been observed before \cite{Deng:1999,Trippenbach:2000}. However, in these experiments the interference patterns are just transitory and only recently has the realisation of the supersolid phase in BECs provided a way to create stationary matter waves with broken, but periodic spatial symmetry. Such phases can be reached by using spin-orbit coupled two-component condensates \cite{Lonard:2017,li_2017} or gases in which the atoms have large dipole moments \cite{stringari_dipolar:2019,ferlaino_dipolar:2019,Guo_2019}. An impurity atom immersed in such a system will experience a periodic mean-field potential, and therefore also the associated band-structure spectrum. However, as the coupling is mediated by scattering, any dynamics of the impurity will feedback onto the condensate density and lead to the excitation of phonon modes in the mean-field lattice which, in turn, can have an effect on the dynamics of the impurity itself \cite{grusdt:2015, grusdt:2017}.

In this work we consider a paradigmatic example of lattice physics by exploring the dynamics of an impurity that is accelerated through the supersolid phase of a spin-orbit coupled BEC. Despite the atomic scattering present in this dynamics, which can lead to the excitation of phonon modes, we show that the impurity is able to undergo Bloch-like oscillations, which can be quite stable and should therefore be observable. In fact, in the weakly interacting regime considered in this work, the main reason for the ultimate decay of the Bloch oscillations is the dispersion of the impurity wavepacket oscillating in the matter wave lattice. We show that this can be mitigated by considering an impurity that has the form of a non-dispersing bright soliton.

\section{Model}
\label{Sec:Model}

We consider an atomic Bose-Einstein condensate of $N$ particles trapped in an external box potential of length $L$. Two different hyperfine states of these atoms represent a pseudo-spin that is used to implement the spin-orbit coupling interaction through a two-photon Raman pulse from two counter-propagating laser beams \cite{spielman_2009}. The system can then be described as a pseudo spin $\frac{1}{2}$, where the standard kinetic term in the Hamiltonian is modified to  account for the coupling of the pseudo-spins to their linear momentum. In the rotating frame of the laser, the single particle Hamiltonian reads (considering one spatial dimension in the direction of the spin-orbit coupling only) \cite{Pitaevskiistringari2016}
\begin{equation}
    H_{SO}=\frac{\hat{p}_x^2}{2m}-\frac{\hbar k_0 \hat{p}_x}{m}\hat{\sigma}_z+V_{{box}}\hat{I}+\frac{\hbar \delta}{2}\hat{\sigma}_z+\frac{\hbar \Omega}{2}
    \hat{\sigma}_x,
\end{equation}
where  ${p}$ and $m$ are the momentum and mass of the atom respectively,  $\Omega$ is the strength of the laser beams which couple the two spin states, $\hat\sigma_i (i=x,y,z)$ are the Pauli spin operators and the external box potential is given by $V_{box}=\{0~\textrm{for}~{-L/2}~\leq~x~\leq~L/2~; ~\infty~\textrm{for}~|x|>L/2\}$. The detuning is  given by $\delta=\Delta \omega_L-\omega_Z$, where $\omega_Z$ is the energy difference between the two pseudo-spin states and $\Delta \omega_L$ is the frequency difference between the counter-propagating laser beams and $k_0$ is the modulus of the wave number difference between the two Raman lasers.

\begin{figure}[tb]
    \centering
    \includegraphics[width=0.9\textwidth]{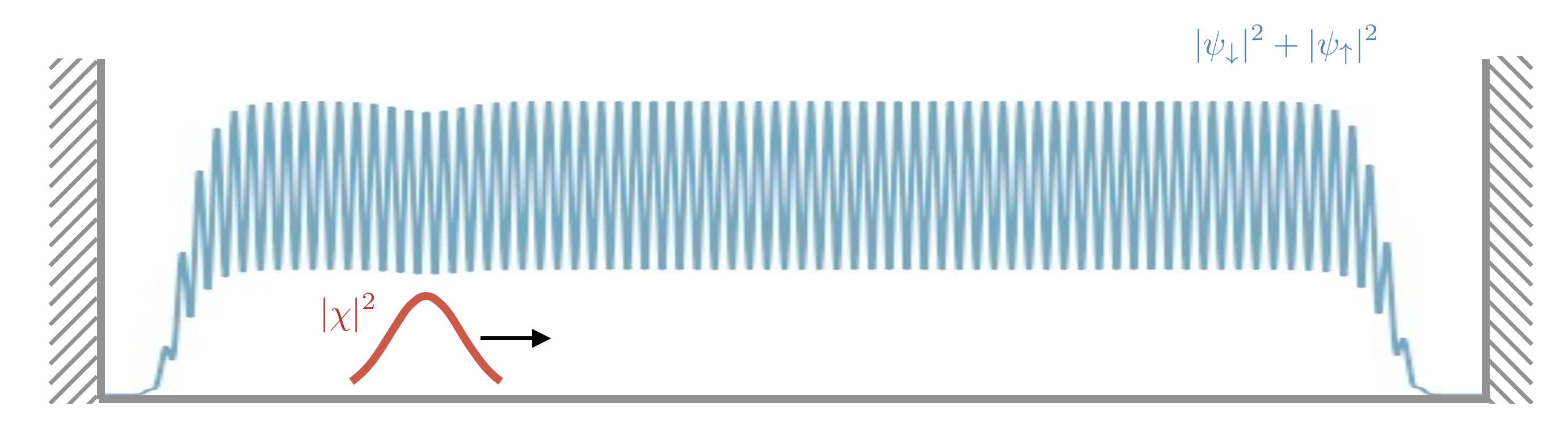}
    \caption{A schematic of the system showing the density of the spin-orbit coupled BEC (blue) and the impurity (red), which is accelerated through the matter-wave lattice.}
    \label{fig:Schematic}
\end{figure}

To consider a many particle system in the mean field limit, one needs to include the spin-dependent interactions, which for  point-like atomic interaction potentials can be quantified by the three parameters $g_{\uparrow\uparrow}, g_{\downarrow\downarrow}$ and $g_{\uparrow\downarrow}=g_{\downarrow\uparrow}$. Taking $\delta=0$ and $g_{\uparrow\uparrow}=g_{\downarrow\downarrow}=g$ for simplicity, the functional for the interaction energy is given by
\begin{eqnarray}
    E_{int}= \frac{1}{4}\int dx\Big[&(g+g_{\uparrow \downarrow})(|\psi_{\uparrow}(x)|^2+|\psi_{\downarrow}(x)|^2)
    \nonumber\\
    &+ (g-g_{\uparrow \downarrow})(|\psi_{\uparrow}(x)|^2-|\psi_{\downarrow}(x)|^2)\Big],
\end{eqnarray}
where $\psi_{\uparrow}$ and $\psi_{\downarrow}$ are the mean-field wavefunctions of the condensate atoms in the spin-up and spin-down states, respectively.
Using the paradigmatic ansatz given by Li \textit{et al.} \cite{li_2012} and minimising the total energy functional in the fully repulsive regime reveals the possibility to reach different ground states depending on the value of the Raman coupling $\Omega$ and the recoil energy of the two photon Raman process given by $E_{rec}=\frac{\hbar^2 k_0^2}{2m}$. In this work we concentrate on the region where $\Omega< 4E_{rec}$ in which the ground state of the condensate is given by a combination of two plane waves with equal and opposite momenta \cite{Pitaevskiistringari2016}. Their interference results in modulations in the density distribution (see Fig.~\ref{fig:Schematic}) and this spontaneous breaking of the continuous translation symmetry, along with the coherent nature of the BEC, defines a supersolid state \cite{li_2017}.

Next we consider an impurity atom immersed into the supersolid phase. 
In the regime where the mean free path of the impurity is of the order of or larger than its de Broglie wavelength, i.e.~in the weakly interacting regime \cite{grusdt:2015,grusdt:2017,Lingua:2018}, the interaction with the condensate can be described by a density-density coupling \cite{Bruderer_2008}, in contrast to the regime in which Bose-polarons were recently reported \cite{parish:2015,parish:2018}. In such a set-up the mean-field density of the BEC acts as an effective potential for the impurity, however any dynamics of the impurity will also have an effect on the condensate density. 
All three components are therefore coupled and the full system can  be described by
\begin{eqnarray}
    \hspace*{-40pt} 
    i\frac{\partial\psi_{\uparrow}}{\partial t} &=\left[-\frac{1}{2}\frac{\partial^2}{\partial x^2}-i \gamma \frac{\partial}{\partial x}+V_{box}(x)+g|\psi_{\uparrow}|^2+ g_{\downarrow \uparrow}|\psi_{\downarrow}|^2+{g}_c|\chi|^2 \right]\psi_{\uparrow}
  + \frac{\Omega}{2}\psi_{\downarrow},
  \label{eq:CGPE1}\\
\hspace*{-40pt} i\frac{\partial\psi_{\downarrow}}{\partial t} &=\left[-\frac{1}{2}\frac{\partial^2}{\partial x^2}+i \gamma \frac{\partial}{\partial x}+V_{box}(x)+g_{\uparrow \downarrow}|\psi_{\uparrow}|^2+g|\psi_{\downarrow}|^2+{g}_c|\chi|^2 \right]\psi_{\downarrow} 
  + \frac{\Omega}{2}\psi_{\uparrow},
  \label{eq:CGPE2}\\
\hspace*{-40pt}  
i\frac{\partial \chi}{\partial t} &=\left[-\frac{1}{2}\frac{\partial^2}{\partial x^2}+ g_c(|\psi_{\uparrow}|^2+|\psi_{\downarrow}|^2)+V_{{ext}}^{{imp}}(x) \right]\chi,
\label{eq:ImpurityDynamics}
\end{eqnarray}

where $\chi$ is the wavefunction  of the impurity atom. To have a more general picture of the ongoing dynamics, we will in the following use dimensionless equations by scaling all energies with respect to the energy scale of the box potential $E_{box}=\hbar^2 /mL^2$ by transforming $x=\frac{x_0}{L}$ and $t = t_0\frac{\hbar}{mL^2}$.
To accelerate the impurity through the condensate, Eq.~(\ref{eq:ImpurityDynamics}) also contains a linear potential $V_{ext}^{imp}$ that is only seen by the impurity. For simplicity we consider the interactions between the impurity atom and the BEC atoms to be species independent and characterised by $g_{c}$. All coupling constants are related to their respective 3D scattering lengths, $a_s^{q}$, by the expression $g'_{q}=\frac{4\pi\hbar^2a_s^{q}}{m}, (q=\uparrow\uparrow,\uparrow\downarrow,\downarrow\uparrow,\downarrow\downarrow,c)$, where $m$ is the mass of the scattering atom and $g_{q}=\frac{g'_{q}}{L E_{{box}}}$. We assume all scattering lengths to be positive, which corresponds to a repulsive interaction between all atoms and components. The rescaled spin-orbit coupling constant is given by $\gamma=\frac{\gamma'}{L E_{box}}$, where $\gamma'=\hbar k_0/m$.
The wavefunctions are normalised as $\int |\psi_{\uparrow}(x,t)|^2 dx=\int |\psi_{\downarrow}(x,t)|^2 dx=\frac{N}{2}$ and $\int |\chi(x,t)|^2 dx=1$. In the following section we will solve Eqs.~(\ref{eq:CGPE1})-(\ref{eq:ImpurityDynamics}) and explore the dynamics of the impurity inside the supersolid phase of spin-orbit coupled BEC for various values of the system parameters.

\section{Bloch Oscillations}
\label{Sec:Results}
To create a well-defined initial state, we assume that the impurity is initially localised at a fixed position inside the supersolid and held by a species selective external harmonic trap of frequency $\omega_x$, with $\pi/\tilde{k}\ll\sqrt{\hbar/(m\omega_x)}\ll L$. Here $\pi/\tilde{k}$ is the {\it lattice constant} of the supersolid, with  $\tilde{k}=\gamma\sqrt{1-\frac{\Omega^2}{4(\gamma^2+\frac{\tilde{n}(g+g_{\uparrow\downarrow})}{4})}}$ and  $\tilde{n}$ denotes the density of the condensate \cite{Pitaevskiistringari2016}. 
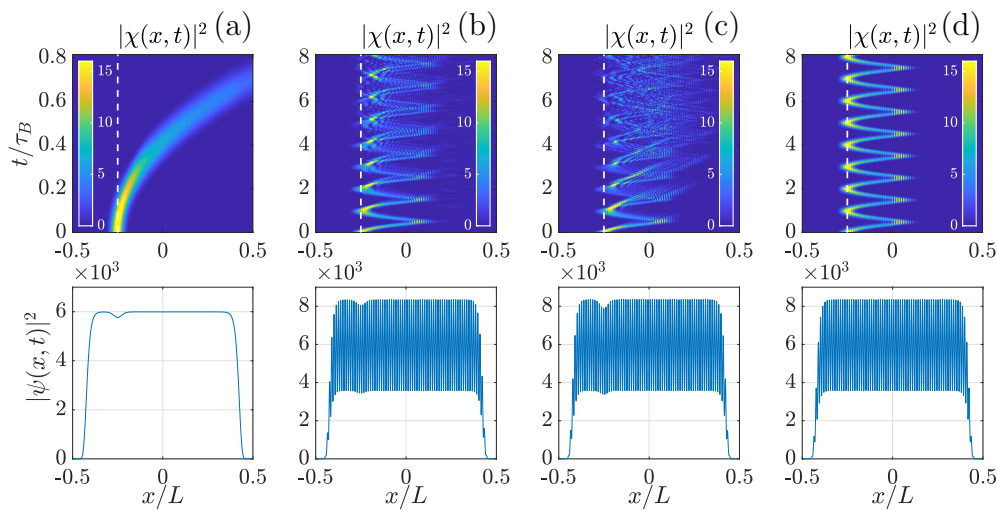
\begin{figure}[tb]
\centering
\begin{subfigure}
\centering
\includegraphics[width=1\textwidth]{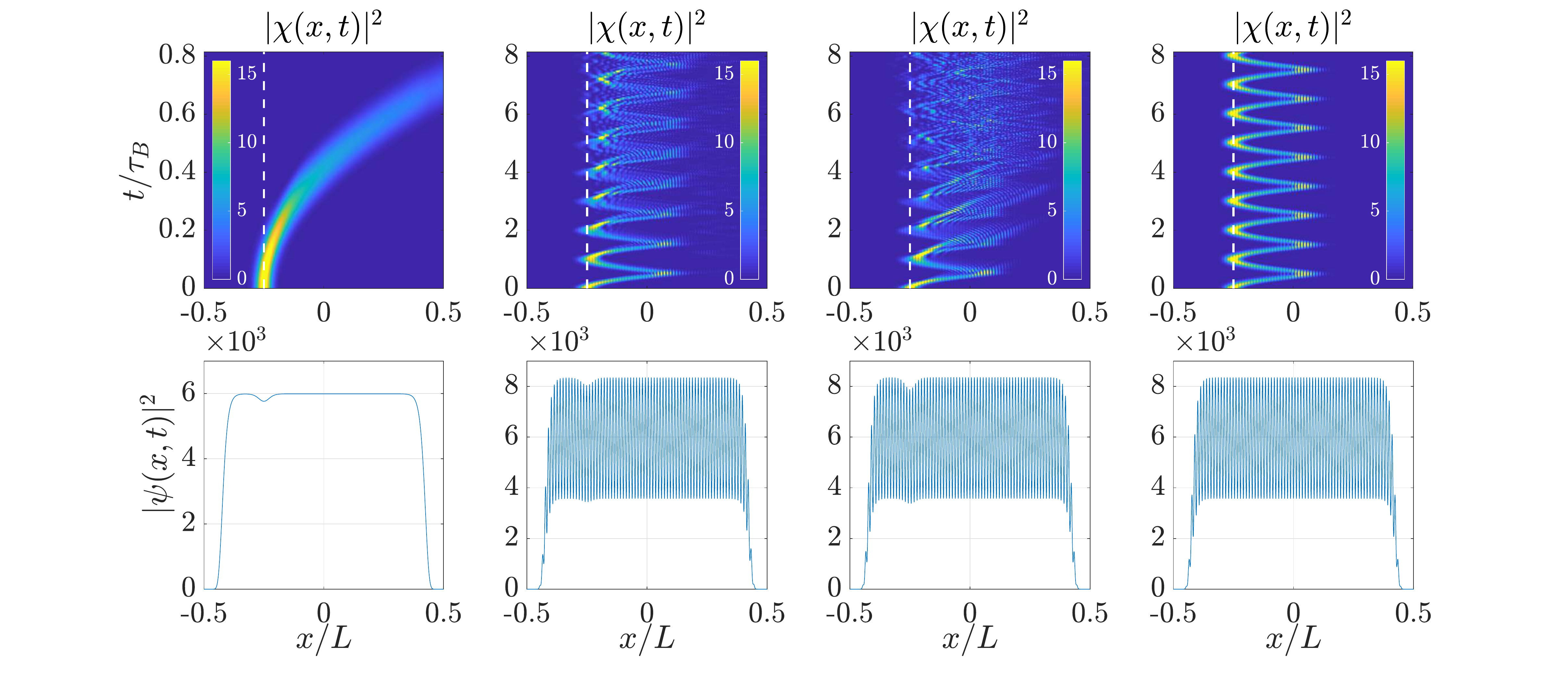}
\end{subfigure}
\put(-335,190){\textcolor{black}{(a)}}
\put(-243,190){\textcolor{black}{(b)}}
\put(-150,190){\textcolor{black}{(c)}}
\put(-60,190){\textcolor{black}{(d)}}
\caption{Time evolution of the density distribution of the impurity (top) and density of the spin-orbit coupled BEC at $t=0$ (bottom) for the three different cases of (a) $\Omega=0$, $\frac{g_c}{L}=3.6$, (b) $\Omega=6.4\times10^4$, $\frac{g_c}{L}=3.6$, (c) $\Omega=6.4\times10^4$, $\frac{g_c}{L}=5.2$ and (d)  in the artificial setting where the lattice is frozen in time leading to non-dispersive oscillations of the impurity with $\Omega=6.4\times10^4$, $\frac{g_c}{L}=5.2$. Other parameters of the system are $\frac{g_{\downarrow \downarrow}}{L}=\frac{g_{\uparrow \uparrow}}{L}$ = $1.6 \times 10^3$ and $\frac{g_{\downarrow \uparrow}}{L}=\frac{g_{\uparrow \downarrow}}{L}=8 \times 10^2$, $N=5\times10^3$, $\frac{\gamma}{L}=2.8\times10^2$ and $\alpha L
=8.32\times10^4$ and the lattice spacing $\frac{d}{L} = 1.23\times10^{-2} $ in the box units.}
\label{fig:densityplots}
\end{figure}
The impurity therefore extends over multiple lattice sites of the supersolid and its energy is larger than the fringe depth of the lattice, cf.~\cite{arimondo:2001}. The coupled ground state of the system is then obtained using a split-step Fourier method and imaginary time propagation \cite{weideman_1986, Bao2003}. Since the interaction between the BEC and the impurity is assumed to be repulsive, the initial state of the BEC has a density dip at the position of the impurity (see lower row in Fig.~\ref{fig:densityplots}) and the density of the impurity is weakly corrugated with the matter wave lattice periodicity. To accelerate the impurity and start the dynamics at $t=0$, the harmonic potential is switched off and replaced by a linear ramp $V_{ext}^{imp}(x,t>0)=-\alpha x$. As a reference case, we first consider the BEC in the absence of the supersolid modulation, $\Omega=0$, when the condensate density is flat. As one would expect, in this situation the acceleration leads to the impurity simply getting expelled from the gas (see Fig.~\ref{fig:densityplots}(a)).

Contrary to this, in the presence of a lattice structure and for small, but finite values of coupling between the impurity and the BEC one can see from Fig.~\ref{fig:densityplots}(b) that the impurity remains trapped within the condensate by undergoing oscillations. These oscillations can be identified as Bloch oscillations, as their period matches closely with the standard textbook prediction for this effect given by  $\tau_B = \frac{2\pi}{\alpha d} = 6.14\times10^{-3}$ in the box units \cite{Kittel:2004}. Here $\alpha$ is the strength of the linear force and $d$ is the lattice spacing. These oscillations are similar to the Bloch oscillations known for electrons in solid state systems \cite{Ashcroft76,Hartmann_2004} or for atoms in optical lattices \cite{arimondo:2001,oberthaler_2006,Anderson:1998,Kessler:2016,prasanna:2009}, but here they are due to the periodic structure of the supersolid matter wave potential. Once the impurity has picked up enough momentum, it gets Bragg reflected at the band edge and starts running up the linear potential again until its approximate starting point. There it gets reflected due to the linear potential and is accelerated through the supersolid again. Let us stress that the numerical simulations are performed on a grid large enough  that boundary effects do not affect the results shown.

However, one can also see from Fig.~\ref{fig:densityplots}(b) that over time these oscillations dephase and that the impurity wavefunction disperses. This is partly due to the initial state at $t=0$ not being an eigenstate of the system (i.e.~once its harmonic oscillator potential is being removed) and partly due to the interaction between the impurity and the BEC. The latter leads to the creation of phonons in the condensate and therefore changes the energy and the momentum distribution of the impurity.  The dephasing of the oscillations and the dispersing of the impurity wavefunction is even faster for increased scattering with the matter wave, see Fig.~\ref{fig:densityplots}(c). For $\frac{g_c}{L}=5.2$ the impurity wavefunction very quickly delocalises across the whole condensate, however most of it is still located in between the two turning points of the oscillation. To show that this dephasing of the impurity wavefunction is due to the scattering with the matter-wave lattice and not due to, say, tunnelling into higher bands, we show in Fig.~\ref{fig:densityplots}(d) the situation where the feedback of the impurity motion onto the condensate is artificially turned off.  As can be seen, in the absence of induced phonon modes in the BEC the impurity oscillates almost perfectly over a long time.

\begin{figure}[tb]
\centering
\begin{subfigure}
\centering
\includegraphics[width=1\textwidth]{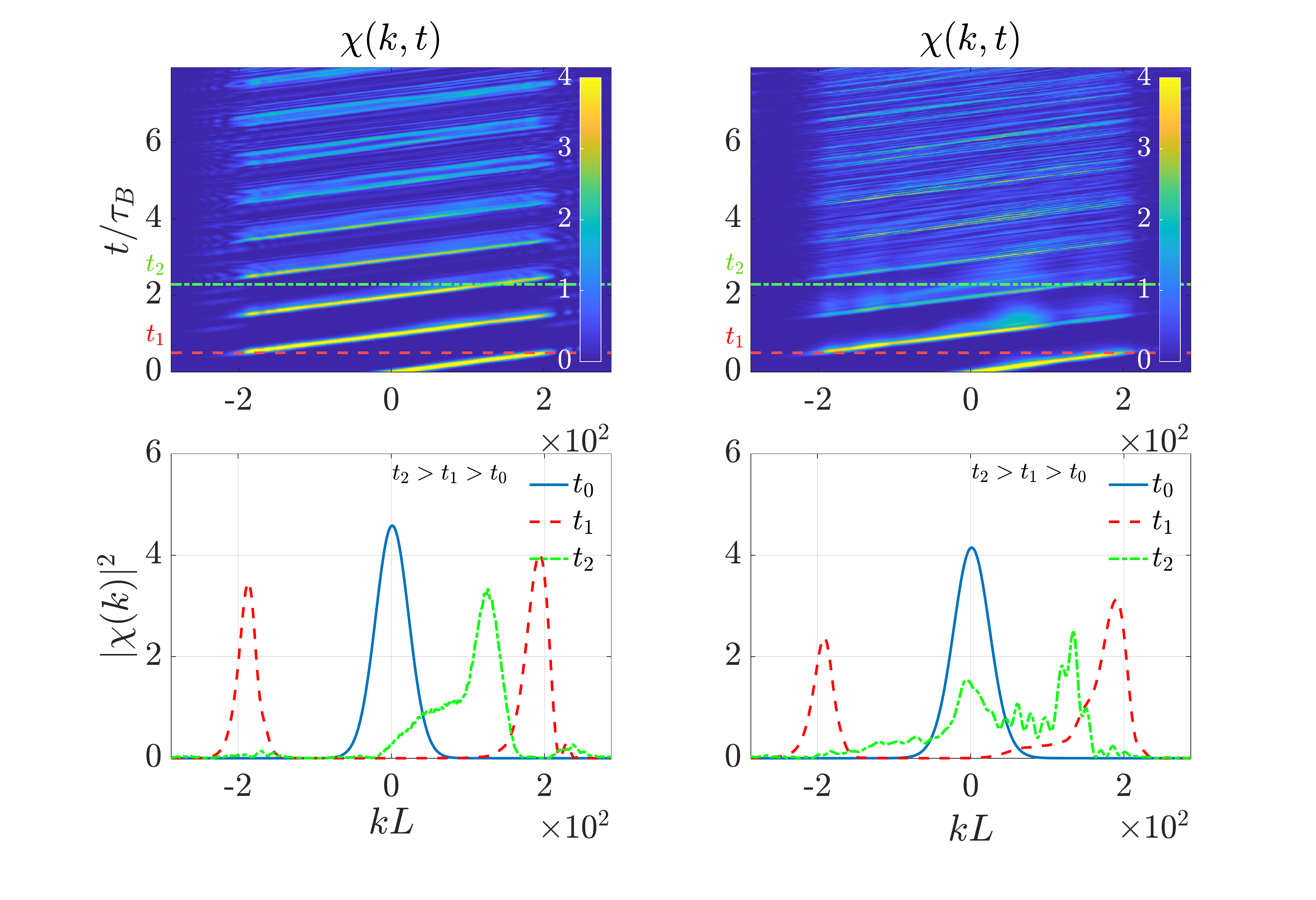}
\end{subfigure}
\put(-260,287){\textcolor{black}{\textbf{(a)}}}
\put(-63,287){\textcolor{black}{\textbf{(b)}}}
\caption{The momentum distribution of the impurity as a function of time (top) corresponding to $\frac{g_c}{L}=3.6$ and $\frac{g_c}{L}=5.2$. The bottom panels show the momentum peaks at three different times, namely $t_0=0$, $t_1 = 0.508\tau_B$ and $t_2= 2.341\tau_B$ in the box units.}
\label{fig:momentumplots}
\end{figure}

The redistribution of momentum of the impurity due to scattering is shown in Fig.~\ref{fig:momentumplots} for the same weak and strong BEC-impurity coupling used in Figs.~\ref{fig:densityplots}(b) and (c). One can see that the momentum distribution, $\chi(k,t) = \frac{1}{2\pi}\int dx\, e^{-ikx}\chi(x,t)$, is initially localised around zero and then evolves due to the linear potential, until it hits the point where it gets Bragg reflected. The momentum for which this happens is, for a plane wave, given by the Bragg reflection criterion, $2d\sin \theta = \lambda$, where $d$ is the lattice spacing. Setting $\sin \theta = 1$ and rewriting $\lambda = \frac{2\pi}{k}$, we have the standard condition $k=\frac{\pi}{d}$, which gives $k\approx 2.6\times10^2$ in box units and which is close to the numerically observed value for our Gaussian wavepacket.

In addition Fig.~\ref{fig:momentumplots} shows that the momentum distribution of the impurity delocalises over time due to scattering with the matter-wave atoms. The top row in Fig.~\ref{fig:momentumplots} shows the momentum distribution as a function of time while the bottom row displays snapshots for the state at the initial time ($t_0=0$), around the time of the first Bragg reflection ($t_1 = 0.497\tau_B$) and at a later stage after two reflections on each side have happened ($t_2= 2.28\tau_B$). One can see that for weaker interactions  the momentum distribution of the impurity starts to de-phase after a few reflections, however it never becomes fully delocalised on the scale displayed. In contrast to this, the momentum distribution for the stronger interacting case $\frac{g_c}{L}=5.2$ very quickly delocalises and while the impurity is mostly trapped within the supersolid, it does no longer behave like one oscillating particle. 

\begin{figure}[tb]          
    \centering
    \includegraphics[width=\textwidth]{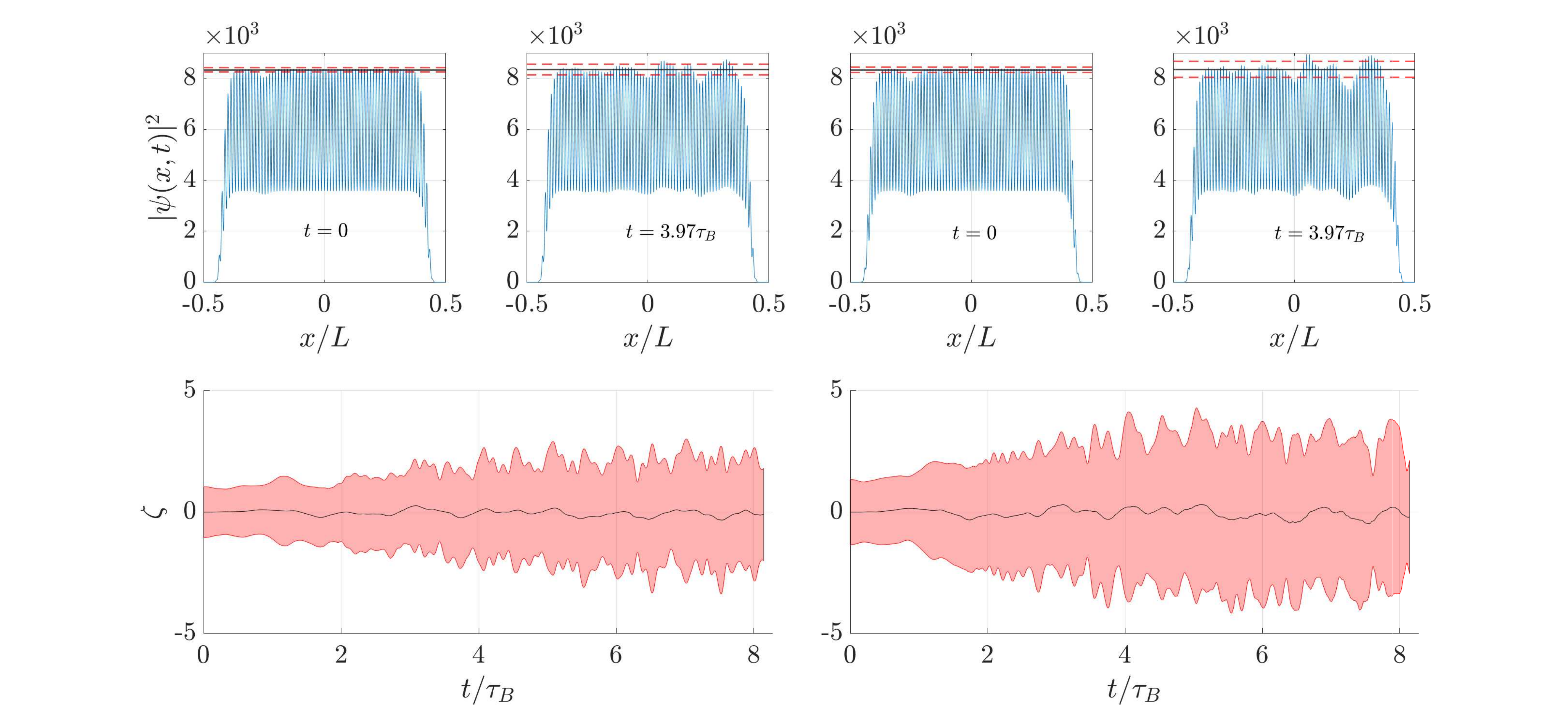}
    \put(-245,195){\textcolor{black}{\textbf{(a)}}}
    \put(-60,195){\textcolor{black}{\textbf{(b)}}}
    \caption{The lattice maxima as a function of time for the cases of $\frac{g_c}{L}=3.6$ and $\frac{g_c}{L}=5.2$ respectively. The upper panels show the condensate density as a function of position at $t_1=0$ and $t_2=3.97\tau_B$ with the average lattice maximum indicated by a black line and the variance by red, dashed lines. The lower panels show the changes in these two quantities over time. The other system parameters are the same as in Fig.~\ref{fig:densityplots}}
\label{fig:fluctuationplots}
\end{figure}

Let us next look at the effect the scattering with the impurity has on the supersolid, by examining the density distribution. The destruction of the initial {\it perfect} lattice for $t>0$ can be quantified by looking at the average of the lattice density maxima, $\overline{\rho}_{max}$, and the corresponding variance, $\pm\sigma_{\overline{\rho}_{max}}$, as a function of time (see Fig.~\ref{fig:fluctuationplots}). The average lattice maximum is defined as the mean value of all local maxima of $\left( |\psi_\uparrow(x,t)|^2+|\psi_\downarrow(x,t)|^2 \right)$ within the central region of the spin-orbit coupled BEC, that is $\frac{x}{L}\in (-1/3,1/3)$, and its variance is $\sigma_{\overline{\rho}_{max}}^2 = \langle \overline{\rho}_{max}^2 \rangle - \langle \overline{\rho}_{max} \rangle^2 $. We highlight the changes over time by plotting $\zeta(t) = \overline{\rho}_{max}(t)  -\overline{\rho}_{max}(t=0)$ in
Fig.~\ref{fig:fluctuationplots} for the two different cases
$\frac{g_c}{L}=3.6$ (a) and $\frac{g_c}{L}=5.2$ (b). The upper panels show the lattice structure at two instants of time, namely at $t_1=0$ and $t_2=3.97\tau_B$, with the horizontal black line indicating the average and the dashed red line the variance. Note that the variance at $t_1=0$ is finite as the lattice density has a dip where the impurity is located.  
One can see that the phonon modes that are excited in both cases are very similar at these low energies and on these timescales, and mostly differ in magnitude. This can also be seen in the detailed plots in the lower row of Fig.~\ref{fig:fluctuationplots}, however a large number of differences in the details are visible as well. While the elementary excitations in spin-orbit coupled supersolid are given by a hybridisation of the low lying branches corresponding to phonon and spin excitations \cite{Saccani_2012}, the symmetric coupling of the impurity to both condensate components (see eqs.~(\ref{eq:CGPE1}) - (\ref{eq:ImpurityDynamics})) prevents the spin mode from being excited \cite{Geier:2021} and all excitations seen in  Fig.~\ref{fig:fluctuationplots} come from the standard phonon branch. It is also worth noting that the movement of the impurity does not lead to any significant variation in the lattice periodicity \textit{i.e.} there are no breathing modes in the system. At the energies available in this system, the scattering processes play a role only in the linear phononic part of the spectrum and while the excitation of the roton mode is possible once the impurity has reached a momentum $\hbar k$, we do not observe any significant sign of it.


Since the decay of observable oscillations can be mainly attributed to the dispersion and the scattering with the matter wave lattice, it is interesting to consider this situation for a non-dispersing impurity wavepacket. This can be realised by assuming that the impurity is a small condensate itself, where an intrinsically attractive interaction leads to a bright soliton state \cite{hulet:2002,shabat:1972,malomed:2011}. Adding this non-linear interaction to the evolution equation (\ref{eq:ImpurityDynamics}) of the impurity leads to
\begin{equation}
    i\frac{\partial\phi}{\partial t} =\left[-\frac{1}{2}\frac{\partial^2}{\partial x^2}+g_c(|\psi_{\downarrow}|^2+\psi_{\uparrow}|^2)+V_{{ext}}^{{imp}}(x)-g_s|\phi|^2 \right]\phi,
\end{equation}
where the normalisation is now given by $\int |\phi(x,t)|^2dx =N_s$. Here $N_s$ is the number of atoms in the soliton component and $g_s>0$ is the strength of the self interaction.

\begin{figure}[tb]
    \centering
    \includegraphics[width=1\textwidth]{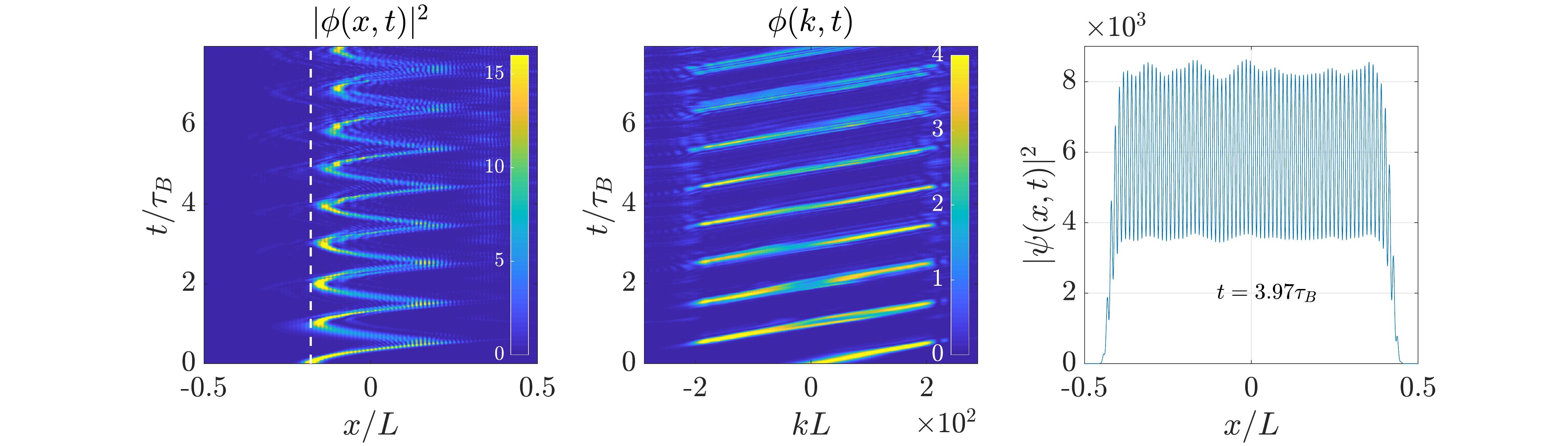}
    \put(-310,120){\textcolor{black}{\textbf{(a)}}}
    \put(-185,120){\textcolor{black}{\textbf{(b)}}}
    \put(-58,120){\textcolor{black}{\textbf{(c)}}}
    \caption{(a) Density of the soliton component as a function of time. One can clearly see that the oscillations are much more pronounced when compared to the single atom case, see Fig.~\ref{fig:densityplots}(b). (b) Momentum distribution of the soliton as a function of time. The momentum does not break into smaller peaks in contrast to the case in \ref{fig:momentumplots}(a). (c) Lattice density showing the phononic amplitude fluctuations at $t=3.97\tau_B$.}
\label{fig:density-momentum-soliton}
\end{figure}

The dynamics in this situation is shown in Fig.~\ref{fig:density-momentum-soliton} for the values $\frac{g_c}{L}=3.6$ and $\frac{g_s}{L}=80$. Comparing to the situation shown in Figs.~\ref{fig:densityplots}(b) and \ref{fig:momentumplots}(a), it is immediately obvious that the impurity undergoes many more well defined oscillations and that the momentum distribution is not spreading out as much. Yet, the soliton still undergoes scattering with the matter wave lattice and loses energy to it, which can be seen from the fact that the upper turning point moves down the linear potential over time. These collisions excite phonon modes in the matter-wave lattice, which leads to amplitude fluctuations only, see Fig.~\ref{fig:density-momentum-soliton}(c). Note that the the ratio of the interaction strengths needs to be $\frac{g_s}{g_c}\gg 1$, in order for the impurity to maintain its solitonic properties. 

\section{Conclusions and Outlook}
In this work we have considered the accelerated motion of a localised atomic impurity that is repulsively coupled to a spin-orbit coupled BEC in the supersolid phase. In this phase the BEC possesses a periodic density distribution, which is experienced by the impurity as a mean-field lattice potential. We have shown that once the momentum of the accelerated impurity fulfils the Bragg condition, it gets reflected and decelerates until coming to a stop at (almost) its initial position again. There the cycle starts over, leading to Bloch-like oscillations.

However, for an initially localised single-atom impurity the momentum distribution disperses over time and the repulsive scattering with the BEC leads to a redistribution of the individual components. Due to both of these effects the initially clearly observable oscillations dephase for longer times and the impurity wavefunction delocalises across the whole mean-field potential. We have shown that this effect can be partly counteracted by considering an impurity that has the form of a bright soliton.

The scattering between the impurity and the BEC also leads to the excitation of phonon modes inside the supersolid. However, due to the low energy of the collisions we consider these only effect the amplitude of the lattice and not disturb the periodicity. In fact the periodicity is determined by the position of the roton-minimum and depends on the details of the spin-orbit coupling term, which requires higher energies to be affected.

The systems we have considered here is an example of a cold atom analogue of a well known condensed matter system. While cold atoms have been used to explore many analogies in the recent two decades using optical lattice settings \cite{Bloch:08}, the system considered here adds the fundamental aspect of scattering with the lattice component and the excitation of phonon modes. 
Even though the experimental implementation of the system considered above is certainly challenging, all ingredients required have been independently demonstrated: Bose-Einstein condensates can be created in flat-bottom box potentials \cite{hadzibabic:2013}, the striped phase in the presence of spin-orbit coupling has been observed \cite{li_2017}, and single impurities can be introduced into condensates in a controlled way \cite{Schmidt:18}. We believe that this system will be a feature-rich addition to the area of quantum simulators in ultracold atom systems with strong connections to condensed matter physics.

\section*{Acknowledgements}
We thank Prof.~P.~Engels, Dr.~K.~Gietka, Dr.~T.~Fogarty and L.~Ruks for fruitful discussions. This work was supported by the Okinawa Institute of Science and Technology Graduate University.
JP also acknowledges the JSPS KAKENHI Grant Number 20K14417. The authors are also grateful for the the Scientific Computing and Data Analysis (SCDA) section of the Research Support Division at OIST.

\newpage
\section*{References}
\bibliographystyle{iopart-num}
\bibliography{references}

\end{document}